%% file: paper338_final.tex
\documentstyle[10pt,epsf,epsfig,hangcaption,xspace,amssymb,amsfonts,amsmath,amsthm,cite,
dp_delphititle]{dp_delphi}
\makeindex
\def\DpPaperGroup{PH-EP}
\def\DpPaperRef{2004-068}
\def\DpDate{03 September 2004}
\def\DpAuthors{DELPHI Collaboration}
\def\DpSubmit{(Accepted by Euro. Phys. Journ. C)}
\def\DpTitle{{ Production of $\Xi{_c^0}$ and $\Xi{_b}$ 
in Z decays and lifetime measurement of $\Xi{_b}$}}
\def\DpComment{  }
\def\DpEMail{  }
\include{newcom}
\newcommand{\jst}{\textsc{Jetset 7.3}\xspace}

\newcommand{\delsim}{\textsc{Delsim}\xspace}
\newcommand{\delana}{\textsc{Delana}\xspace}

\newcommand{\bb}{{{b}\overline{{b}}}\xspace}

\newcommand{\qq}{{{q}\overline{{q}}}\xspace}

\newcommand{\xin}{\mbox{$\Xi^-$}}

\newcommand{\GeVcc} {\mbox{$ {\mathrm{GeV}}/c^2 $}}

\newcommand{\MeVcc} {\mbox{$ {\mathrm{MeV}}/c^2 $}}
\newcommand{\MeVc} {\mbox{$ {\mathrm{MeV}}/c $}}

\newcommand{\Xzero}{\mbox{$\Xi{_c^0}$}}
\newcommand{\Xb}{\mbox{$\Xi{_b}$}}
\newcommand{\Xfem}{\mbox{$\Xi(1530)^0$}}

\begin{document}
\makeatletter
\newcount\@tempcntc
\def\@citex[#1]#2{\if@filesw\immediate\write\@auxout{\string\citation{#2}}\fi
  \@tempcnta\z@\@tempcntb\m@ne\def\@citea{}\@cite{\@for\@citeb:=#2\do
    {\@ifundefined
       {b@\@citeb}{\@citeo\@tempcntb\m@ne\@citea\def\@citea{,}{\bf ?}\@warning
       {Citation `\@citeb' on page \thepage \space undefined}}%
    {\setbox\z@\hbox{\global\@tempcntc0\csname b@\@citeb\endcsname\relax}%
     \ifnum\@tempcntc=\z@ \@citeo\@tempcntb\m@ne
       \@citea\def\@citea{,}\hbox{\csname b@\@citeb\endcsname}%
     \else
      \advance\@tempcntb\@ne
      \ifnum\@tempcntb=\@tempcntc
      \else\advance\@tempcntb\m@ne\@citeo
      \@tempcnta\@tempcntc\@tempcntb\@tempcntc\fi\fi}}\@citeo}{#1}}
\def\@citeo{\ifnum\@tempcnta>\@tempcntb\else\@citea\def\@citea{,}%
  \ifnum\@tempcnta=\@tempcntb\the\@tempcnta\else
   {\advance\@tempcnta\@ne\ifnum\@tempcnta=\@tempcntb \else \def\@citea{--}\fi
    \advance\@tempcnta\m@ne\the\@tempcnta\@citea\the\@tempcntb}\fi\fi}
 
\makeatother
\begin{titlepage}
\pagenumbering{roman}
\CERNpreprint{\DpPaperGroup}{\DpPaperRef} 
\date{{\small\DpDate}} 
\title{\DpTitle} 
\address{\DpAuthors} 
\begin{shortabs} 
\noindent
The charmed strange baryon $\Xi{_c^0}$ was searched for in the decay
channel $\Xi{_c^0}
\rightarrow \Xi^- \pi^+$, and the beauty strange baryon $\Xi{_b}$ in the
inclusive
channel $\Xi_b \rightarrow \Xi^{-} \ell^{-} \bar{\nu} X$,
using the 3.5 million hadronic Z events collected by the DELPHI experiment
in the years 1992--1995.
The $\Xi^-$ was reconstructed through the decay 
$\Xi^- \rightarrow \Lambda \pi^-$, using
a constrained fit method for cascade decays. An iterative
discriminant analysis was used for the $\Xi{_c^0}$ and $\Xi{_b}$ selection.
The production rates were measured to be
$f_{\Xi{_c^0}} \times$BR$(\Xi{_c^0}
\rightarrow \Xi^- \pi^+)=(4.7 \pm 1.4 (stat.) \pm 1.1 (syst.))\times 10^{-4}$
per hadronic Z decay, and
BR$(b \rightarrow \Xi{_b}) \times$BR$(\Xi{_b} \rightarrow
\Xi^- \ell^- X)=(3.0 \pm 1.0(stat.) \pm 0.3(syst.))\times 10^{-4}$
for each lepton species (electron or muon).
The lifetime of the $\Xi{_b}$ baryon was measured to be 
$ \tau_{\Xi{_b}} = 1.45{^{+0.55}_{-0.43}} (stat.) \pm 0.13 (syst.)$~\mbox{ps}.
A combination with the previous DELPHI lifetime measurement gives
$ \tau_{\Xi{_b}} = 1.48{^{+0.40}_{-0.31}} (stat.) \pm 0.12 (syst.)$~\mbox{ps}.

\end{shortabs}
\vfill
\begin{center}
\DpSubmit \ \\ 
\DpComment \ \\
\DpEMail \ \\
\end{center}
\vfill
\clearpage
\headsep 10.0pt
\addtolength{\textheight}{10mm}
\addtolength{\footskip}{-5mm}
\begingroup
%
\newcommand{\DpName}[2]{\hbox{#1$^{\ref{#2}}$},\hfill}
\newcommand{\DpNameTwo}[3]{\hbox{#1$^{\ref{#2},\ref{#3}}$},\hfill}
\newcommand{\DpNameThree}[4]{\hbox{#1$^{\ref{#2},\ref{#3},\ref{#4}}$},\hfill}
\newskip\Bigfill \Bigfill = 0pt plus 1000fill
\newcommand{\DpNameLast}[2]{\hbox{#1$^{\ref{#2}}$}\hspace{\Bigfill}}
%
\footnotesize
\noindent
\DpName{J.Abdallah}{LPNHE}
\DpName{P.Abreu}{LIP}
\DpName{W.Adam}{VIENNA}
\DpName{P.Adzic}{DEMOKRITOS}
\DpName{T.Albrecht}{KARLSRUHE}
\DpName{T.Alderweireld}{AIM}
\DpName{R.Alemany-Fernandez}{CERN}
\DpName{T.Allmendinger}{KARLSRUHE}
\DpName{P.P.Allport}{LIVERPOOL}
\DpName{U.Amaldi}{MILANO2}
% ...
\endgroup
\titlefoot{Department of Physics and Astronomy, Iowa State
     University, Ames IA 50011-3160, USA
    \label{AMES}}
\titlefoot{Physics Department, Universiteit Antwerpen,
     Universiteitsplein 1, B-2610 Antwerpen, Belgium \\
     \indent~~and IIHE, ULB-VUB,
     Pleinlaan 2, B-1050 Brussels, Belgium \\
     \indent~~and Facult\'e des Sciences,
     Univ. de l'Etat Mons, Av. Maistriau 19, B-7000 Mons, Belgium
    \label{AIM}}
\titlefoot{Physics Laboratory, University of Athens, Solonos Str.
     104, GR-10680 Athens, Greece
    \label{ATHENS}}
\titlefoot{Department of Physics, University of Bergen,
     All\'egaten 55, NO-5007 Bergen, Norway
    \label{BERGEN}}
\titlefoot{Dipartimento di Fisica, Universit\`a di Bologna and INFN,
     Via Irnerio 46, IT-40126 Bologna, Italy
    \label{BOLOGNA}}
\titlefoot{Centro Brasileiro de Pesquisas F\'{\i}sicas, rua Xavier Sigaud 150,
     BR-22290 Rio de Janeiro, Brazil \\
     \indent~~and Depto. de F\'{\i}sica, Pont. Univ. Cat\'olica,
     C.P. 38071 BR-22453 Rio de Janeiro, Brazil \\
     \indent~~and Inst. de F\'{\i}sica, Univ. Estadual do Rio de Janeiro,
     rua S\~{a}o Francisco Xavier 524, Rio de Janeiro, Brazil
    \label{BRASIL}}
\titlefoot{Coll\`ege de France, Lab. de Physique Corpusculaire, IN2P3-CNRS,
     FR-75231 Paris Cedex 05, France
    \label{CDF}}
\titlefoot{CERN, CH-1211 Geneva 23, Switzerland
    \label{CERN}}
\titlefoot{Institut de Recherches Subatomiques, IN2P3 - CNRS/ULP - BP20,
     FR-67037 Strasbourg Cedex, France
    \label{CRN}}
\titlefoot{Now at DESY-Zeuthen, Platanenallee 6, D-15735 Zeuthen, Germany
    \label{DESY}}
\titlefoot{Institute of Nuclear Physics, N.C.S.R. Demokritos,
     P.O. Box 60228, GR-15310 Athens, Greece
    \label{DEMOKRITOS}}
\titlefoot{FZU, Inst. of Phys. of the C.A.S. High Energy Physics Division,
     Na Slovance 2, CZ-180 40, Praha 8, Czech Republic
    \label{FZU}}
\titlefoot{Dipartimento di Fisica, Universit\`a di Genova and INFN,
     Via Dodecaneso 33, IT-16146 Genova, Italy
    \label{GENOVA}}
\titlefoot{Institut des Sciences Nucl\'eaires, IN2P3-CNRS, Universit\'e
     de Grenoble 1, FR-38026 Grenoble Cedex, France
    \label{GRENOBLE}}
\titlefoot{Helsinki Institute of Physics and Department of Physical Sciences,
     P.O. Box 64, FIN-00014 University of Helsinki, 
     \indent~~Finland
    \label{HELSINKI}}
\titlefoot{Joint Institute for Nuclear Research, Dubna, Head Post
     Office, P.O. Box 79, RU-101 000 Moscow, Russian Federation
    \label{JINR}}
\titlefoot{Institut f\"ur Experimentelle Kernphysik,
     Universit\"at Karlsruhe, Postfach 6980, DE-76128 Karlsruhe,
     Germany
    \label{KARLSRUHE}}
\titlefoot{Institute of Nuclear Physics PAN,Ul. Radzikowskiego 152,
     PL-31142 Krakow, Poland
    \label{KRAKOW1}}
\titlefoot{Faculty of Physics and Nuclear Techniques, University of Mining
     and Metallurgy, PL-30055 Krakow, Poland
    \label{KRAKOW2}}
\titlefoot{Universit\'e de Paris-Sud, Lab. de l'Acc\'el\'erateur
     Lin\'eaire, IN2P3-CNRS, B\^{a}t. 200, FR-91405 Orsay Cedex, France
    \label{LAL}}
\titlefoot{School of Physics and Chemistry, University of Lancaster,
     Lancaster LA1 4YB, UK
    \label{LANCASTER}}
\titlefoot{LIP, IST, FCUL - Av. Elias Garcia, 14-$1^{o}$,
     PT-1000 Lisboa Codex, Portugal
    \label{LIP}}
\titlefoot{Department of Physics, University of Liverpool, P.O.
     Box 147, Liverpool L69 3BX, UK
    \label{LIVERPOOL}}
\titlefoot{Dept. of Physics and Astronomy, Kelvin Building,
     University of Glasgow, Glasgow G12 8QQ
    \label{GLASGOW}}
\titlefoot{LPNHE, IN2P3-CNRS, Univ.~Paris VI et VII, Tour 33 (RdC),
     4 place Jussieu, FR-75252 Paris Cedex 05, France
    \label{LPNHE}}
\titlefoot{Department of Physics, University of Lund,
     S\"olvegatan 14, SE-223 63 Lund, Sweden
    \label{LUND}}
\titlefoot{Universit\'e Claude Bernard de Lyon, IPNL, IN2P3-CNRS,
     FR-69622 Villeurbanne Cedex, France
    \label{LYON}}
\titlefoot{Dipartimento di Fisica, Universit\`a di Milano and INFN-MILANO,
     Via Celoria 16, IT-20133 Milan, Italy
    \label{MILANO}}
\titlefoot{Dipartimento di Fisica, Univ. di Milano-Bicocca and
     INFN-MILANO, Piazza della Scienza 2, IT-20126 Milan, Italy
    \label{MILANO2}}
\titlefoot{IPNP of MFF, Charles Univ., Areal MFF,
     V Holesovickach 2, CZ-180 00, Praha 8, Czech Republic
    \label{NC}}
\titlefoot{NIKHEF, Postbus 41882, NL-1009 DB
     Amsterdam, The Netherlands
    \label{NIKHEF}}
\titlefoot{National Technical University, Physics Department,
     Zografou Campus, GR-15773 Athens, Greece
    \label{NTU-ATHENS}}
\titlefoot{Physics Department, University of Oslo, Blindern,
     NO-0316 Oslo, Norway
    \label{OSLO}}
\titlefoot{Dpto. Fisica, Univ. Oviedo, Avda. Calvo Sotelo
     s/n, ES-33007 Oviedo, Spain
    \label{OVIEDO}}
\titlefoot{Department of Physics, University of Oxford,
     Keble Road, Oxford OX1 3RH, UK
    \label{OXFORD}}
\titlefoot{Dipartimento di Fisica, Universit\`a di Padova and
     INFN, Via Marzolo 8, IT-35131 Padua, Italy
    \label{PADOVA}}
\titlefoot{Rutherford Appleton Laboratory, Chilton, Didcot
     OX11 OQX, UK
    \label{RAL}}
\titlefoot{Dipartimento di Fisica, Universit\`a di Roma II and
     INFN, Tor Vergata, IT-00173 Rome, Italy
    \label{ROMA2}}
\titlefoot{Dipartimento di Fisica, Universit\`a di Roma III and
     INFN, Via della Vasca Navale 84, IT-00146 Rome, Italy
    \label{ROMA3}}
\titlefoot{DAPNIA/Service de Physique des Particules,
     CEA-Saclay, FR-91191 Gif-sur-Yvette Cedex, France
    \label{SACLAY}}
\titlefoot{Instituto de Fisica de Cantabria (CSIC-UC), Avda.
     los Castros s/n, ES-39006 Santander, Spain
    \label{SANTANDER}}
\titlefoot{Inst. for High Energy Physics, Serpukov
     P.O. Box 35, Protvino, (Moscow Region), Russian Federation
    \label{SERPUKHOV}}
\titlefoot{J. Stefan Institute, Jamova 39, SI-1000 Ljubljana, Slovenia
     and Laboratory for Astroparticle Physics,\\
     \indent~~Nova Gorica Polytechnic, Kostanjeviska 16a, SI-5000 Nova Gorica, Slovenia, \\
     \indent~~and Department of Physics, University of Ljubljana,
     SI-1000 Ljubljana, Slovenia
    \label{SLOVENIJA}}
\titlefoot{Fysikum, Stockholm University,
     Box 6730, SE-113 85 Stockholm, Sweden
    \label{STOCKHOLM}}
\titlefoot{Dipartimento di Fisica Sperimentale, Universit\`a di
     Torino and INFN, Via P. Giuria 1, IT-10125 Turin, Italy
    \label{TORINO}}
\titlefoot{INFN,Sezione di Torino and Dipartimento di Fisica Teorica,
     Universit\`a di Torino, Via Giuria 1,
     IT-10125 Turin, Italy
    \label{TORINOTH}}
\titlefoot{Dipartimento di Fisica, Universit\`a di Trieste and
     INFN, Via A. Valerio 2, IT-34127 Trieste, Italy \\
     \indent~~and Istituto di Fisica, Universit\`a di Udine,
     IT-33100 Udine, Italy
    \label{TU}}
\titlefoot{Univ. Federal do Rio de Janeiro, C.P. 68528
     Cidade Univ., Ilha do Fund\~ao
     BR-21945-970 Rio de Janeiro, Brazil
    \label{UFRJ}}
\titlefoot{Department of Radiation Sciences, University of
     Uppsala, P.O. Box 535, SE-751 21 Uppsala, Sweden
    \label{UPPSALA}}
\titlefoot{IFIC, Valencia-CSIC, and D.F.A.M.N., U. de Valencia,
     Avda. Dr. Moliner 50, ES-46100 Burjassot (Valencia), Spain
    \label{VALENCIA}}
\titlefoot{Institut f\"ur Hochenergiephysik, \"Osterr. Akad.
     d. Wissensch., Nikolsdorfergasse 18, AT-1050 Vienna, Austria
    \label{VIENNA}}
\titlefoot{Inst. Nuclear Studies and University of Warsaw, Ul.
     Hoza 69, PL-00681 Warsaw, Poland
    \label{WARSZAWA}}
\titlefoot{Fachbereich Physik, University of Wuppertal, Postfach
     100 127, DE-42097 Wuppertal, Germany \\
\noindent
{$^\dagger$~deceased}
    \label{WUPPERTAL}}
\normalsize
\addtolength{\textheight}{-10mm}
\addtolength{\footskip}{5mm}
\clearpage
\headsep 30.0pt
\end{titlepage}
%
\pagenumbering{arabic} 
\setcounter{footnote}{0} %
\large

\section{Introduction}

Measuring the production rates of baryons, and heavy baryons in
particular, is important in order to understand the underlying
fragmentation process in $Z \rightarrow \qq$ events.
The fragmentation process involves small momentum transfers and
perturbation theory is not applicable, consequently no good
theoretical description exists and phenomenolo{\mbox gi}cal models 
have to be used.
The production of a baryon-pair requires the creation of a
di-quark pair in the fragmentation. The exact nature of the mechanism
by which this occurs is still largely unknown.
Thus the tuning of fragmentation models and the understanding of the 
processes involved in baryon production, require good measurements of the
production of all the baryons in general, and baryons containing
heavy quarks in particular.

 In this paper, a first measurement of the production at the $Z$ resonance
of the charm strange baryon $\Xi{_c^0}$ is presented
\footnote{Charge conjugated states are implied
throughout this paper, unless otherwise stated.}, using
the exclusive decay channel $\Xi{_c^0} \rightarrow \Xi^- \pi^+$.
As a cross-check, the $\Xi(1530)^0$ resonance is reconstructed,
through the decay channel $\Xi(1530)^0 \rightarrow \Xi^- \pi^+$.

A measurement of the production and lifetime
of the strange $b$-baryon $\Xi_b$ 
is also presented, using the semileptonic 
decay channel, $\Xi_b \rightarrow \Xi^{-} \ell^{-} \bar{\nu} X$.
Here $\Xi_b$ is used as a notation for the strange $b$-baryon states
$\Xi_b^{-}$ and $\Xi_b^{0}$. The $\Xi_b$ baryon will decay to
$X_c X \ell^{-} \bar{\nu}$
followed by $X_c \rightarrow \Xi^{-} X^{'}$,
where $X_c$ is a charmed baryon which yields a $\Xi^-$ hyperon.
$X_c$ is dominantly $\Xi_c^{0}$,
thus the most common state in this decay channel is the $\Xi_b^{-}$.

A first observation of the $\Xi_b$ baryon production and lifetime
has been published by DELPHI, using a smaller data sample~\cite{xibdelphi},
and has been confirmed by ALEPH~\cite{xibaleph}.
Here the full LEP1 data sample collected by the DELPHI experiment
between the years 1992--1995 is used.
In this paper, the background is determined from the data whereas 
simulation was used in the previous DELPHI analysis.
The new lifetime measurement is statistically independent from the
previous DELPHI measurement and relies on a different method.

A constrained multivertex fit has been performed to
reconstruct the $\Xi^- \rightarrow \Lambda \pi^-$decay.
For the $\Xi{_c^0}$, $\Xi(1530)^0$ and $\Xi_b$ selections,
an iterative discriminant analysis method has been applied.

\section{The apparatus}

The DELPHI detector is described in detail 
elsewhere~\cite{Delphi:detector,Delphi:performance}.
The detectors most important for this analysis 
are the Vertex Detector (VD), the Inner Detector (ID),
the Time Projection Chamber (TPC), and the Outer Detector (OD).
For the lepton identification in the $\Xi_b$ analysis
the electromagnetic calorimeter (HPC) and the muon chambers were 
also used.

The VD consists of three concentric layers of silicon-strip detectors,
located at radii of 6~cm, 9~cm and 11~cm from the beam axis.
The polar angles 
\footnote{In the standard DELPHI coordinate system, 
the $z$ axis is along the electron beam direction, the $x$ axis 
points towards the center of LEP, and the $y$ axis points upwards. 
The polar angle to the $z$ axis is called $\theta$ and the azimuthal 
angle around the $z$ axis is called $\phi$; 
the radial coordinate is $R = \sqrt{x^2+y^2}$.}
covered for particles crossing all three layers are
$44 ^\circ < \theta < 136 ^\circ$. 
In 1994 and 1995, the first and third layers of the VD
had double-sided readout and gave both $R\phi$ and $z$ coordinates.
The TPC is the main tracking device where charged-particle tracks are 
reconstructed in three dimensions for radii between 40~cm and 110~cm.
The ID and OD are two drift chambers located at radii between 12~cm and 28~cm
and between 198~cm and 206~cm, respectively, and
provide additional points for the track reconstruction.

The $b$-tagging package developed by the 
DELPHI collaboration~\cite{Delphi:performance,btag} 
has been used to select $Z \rightarrow \bb $ events.
The impact parameters of the charged-particle tracks, with respect to
the primary vertex, have been used to build the probability that all tracks
come from the primary vertex. Due to the long $b$-hadron lifetime, the
probability distribution is peaked at zero for events containing
$b$-quarks whereas it is flat for events containing light quarks only.

\section{Event selection and simulation}
\label{sec:eventsimu}

Hadronic Z decays were selected by requiring at 
least four reconstructed charged particles and a total energy 
of these particles (assumed to be pions) larger than
12\% of the centre-of-mass energy.
The charged-particle tracks had to be longer than 30~cm in the TPC,
with a momentum larger than 400~\MeVc\ 
and a polar angle between 20$^\circ$ and 160$^\circ$.
The polar angle of the thrust axis, $\theta_{thrust}$, was computed for each event
and events were rejected if $|\cos \theta_{thrust}|$ was greater than
0.95. With these requirements the efficiency for the hadronic Z selection
was larger than 95\%.
A total of 3.5 million hadronic events were selected.

Simulated events were produced by the \jst parton-shower
generator~\cite{JETSETfys} and then processed through a detailed simulation
program, \delsim, which modelled the detector response~\cite{Delphi:performance}. 
The simulated result from \delsim was then processed by
the same reconstruction program as used for the data, 
\delana~\cite{Delphi:performance}. 
A total of 9.8 million Z$\rightarrow \qq$ events was
simulated (11.8 million for the $\Xi_b$ analysis).

In this simulation sample, there were only a few thousand  
$\Xi{_c^0} \rightarrow \Xi^- \pi^+$ events,
and about 1000 $\Xi_b \rightarrow \Xi \ell \bar{\nu} X$ events.
Thus some dedicated $\Xi{_c^0}$ and $\Xi_b$ samples were generated for the 
years 1992--1995 using the \delsim and \delana versions corresponding 
to each year. 
About \mbox{58 000} $\Xi{_c^0} \rightarrow \Xi^- \pi^+$ events
and \mbox{208 000} $\Xi_b \rightarrow \Xi \ell \bar{\nu} X$
events were simulated (see Table~\ref{nrofxicb}).
\begin{table}[htb]
\begin{center}
\begin{tabular}{|l|r|r|r|}\hline
Year  & $\Xi{_c^0}$ events  & $\Xi_b^{-}$ events & $\Xi_b^{0}$  events\\
\hline
\hline
1992  &  11 159   &  50 710 & 1 365 \\
1993  &  12 142   &  52 735 & 1 331 \\
1994  &  20 636   &  50 203 & 1 221 \\
1995  &  13 874   &  49 378 & 1 227 \\
\hline
\hline
Total &  57 811   & 203 026 & 5 144 \\
\hline
\end{tabular}
\protect\caption{ \label{nrofxicb} {\em Number 
of simulated\/ $\Xi{_c^0} \rightarrow \Xi^- \pi^+$, $\Xi_b^{-}$ 
and $\Xi_b^{0}$ events for each 1992--1995 year.}}
\end{center}
\end{table}

\section{$\Xi{^-}$ reconstruction}
\label{sec:Xirec}

The $\Xi^-$ hyperon was
reconstructed through the decay $\Xi^- \rightarrow \Lambda \pi^-$.
A constrained multivertex fit to the three-dimensional decay topology
was used to reconstruct the decay chain and suppress the combinatorial
background.

In this analysis all V$^0$ candidates, i.e.~all pairs of oppositely
charged particles, were considered as $\Lambda$ candidates. 
For each pair, the highest momentum particle was assumed to be a proton and 
the other a pion, and a vertex fit was performed
by the standard DELPHI $\rm V^{0}$ search algorithm~\cite{Delphi:performance}.
The $\Lambda$ candidates were selected by requiring:
an invariant mass $M(p \pi^-)$ between 1.107~GeV$/c^2$ and 1.125~GeV$/c^2$, 
a $\chi^2$ probability of the V$^0$ vertex fit larger than 0.001 
and an $R \phi$ decay length greater than 1.0~cm.
To avoid gamma conversions, the relative transverse momentum, $p_T$,
of the proton and pion had to be greater than 0.03 GeV$/c$, 
with $p_T$ calculated with respect to the line joining the primary and
 secondary vertices. The angle in the $xy$-plane between the V$^0$ 
momentum and its line of flight had to be
smaller than 0.08 radian. If the V$^0$ was reconstructed outside the VD, 
it was also required that no signal in the vertex detector could 
be consistently associated with the V$^0$ vertex tracks.
In the following, the proton and pion from the $\Lambda$ decay will be called
$p_1$ and $\pi_1$, respectively, thus denoting $\Lambda=(p{_1}\pi{_1^-})$.

The $\Lambda$'s selected as described above were combined with pions
(called $\pi_2$ in the following) with the same charge as the $\pi_1$ from
the $\Lambda$.
The $\pi_2$ candidate had not to be tagged as an electron or a muon.

A constrained multivertex fit~\cite{omsig,blobel}
was performed if the invariant mass $M(\Lambda \pi{_2^-})$ was
smaller than 2.0~GeV$/c^2$ and if
the distance between the two trajectories in the $z$ direction was 
smaller than $2$ cm at the point of crossing in the $xy$-plane.
The fit used was a general least-squares fit with the following kinematical 
and geometrical constraints applied to each \xin\ candidate:
\begin{itemize}
\item [-] {the invariant mass $M(p{_1} \pi{_1^-})$ had to be equal
  to the nominal mass of the $\Lambda$;}

\item [-] {the $R \phi$ and $z$ coordinates of $p_1$ and $\pi_1$
  had to be the same at the radial distance of the $\Lambda$ decay point;}

\item [-] {the $R \phi$ and $z$ coordinates of $\Lambda$ and $\pi_2$ 
  had to be equal at the radial distance of the $\Xi^-$ decay point;}

\item [-] {to ensure momentum conservation at the $\Xi^-$ decay point, 
  the polar and azimuthal angles of 
  the $\Xi^-$ candidate had to be equal to the angles from the 
  curved trajectory between the decay and primary vertices. 
  The curvature was calculated from the $\Xi^-$ momentum.}
\end{itemize}

For each of the three particles, i.e.~the proton and pion from the
$\Lambda$ and the pion from the $\Xi^-$, the fit was performed
with the following five track parameters: $1/r$ ($r$ being the radius of
curvature of the track), 
the $z$ and $R\phi$ impact parameters, the polar angle $\theta$, and
the azimuthal angle $\phi$. These 15 variables, plus the $z$ coordinate of the
primary vertex, were the measured variables in the fit. 
The $x$ and $y$ coordinates of the primary vertex 
were so precisely measured that they were taken as fixed.
The unmeasured variables were the radial distances of the
$\Xi^-$ and $\Lambda$ decay points, giving a total of 18
variables in the fit.
The curved \xin\ track was not measured, but calculated in the fit.
All the tracks were corrected for ionization losses, according
to the given mass hypothesis.
The performance of the fit was tuned by adjusting the covariance
matrices of the tracks. The adjustment consisted in
a scaling of the errors of the track parameters.
After the adjustment the pull distributions of the 16 fitted
quantities were standard normal distributions within 10\%.

\begin{figure}[htb]
\begin{center}
\mbox{\epsfig{figure=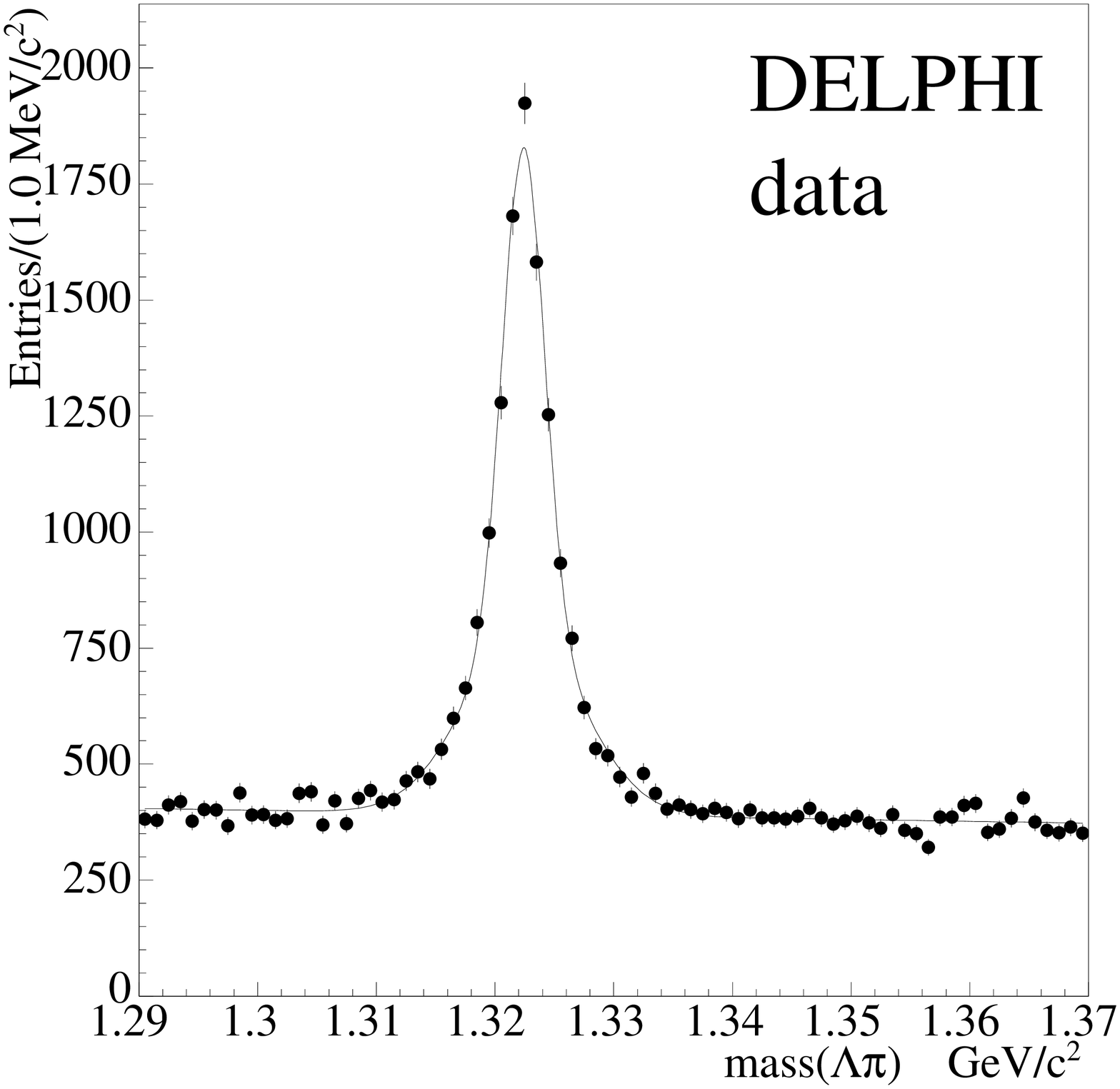,width=10cm}}
\protect\caption[mXi]{\label{mXi}  
{\em The $\Lambda \pi{^-}$ 
invariant-mass spectrum,
using a constrained fit on the 1992--1995 data sample.
The curve is the result of 
a fit using a first-order polynomial to parametrize the background and 
two Gaussian functions of same mean for the signal.
}}
\end{center}
\end{figure}

The events for which the fit converged gave
the $\Lambda \pi{_2^-}$ invariant-mass spectrum 
shown in Figure~\ref{mXi}, 
The solid line is a fit, for illustrative purpose, using two
Gaussian functions with the same mean for the signal, 
and a first order polynomial for
the background, yielding 9445$\pm$584 $\Xi^-$.

\section{Iterative discriminant analysis}
\label{sec:IDA}

Using sequential cuts to select the signal leads to what can be
pictured as a multi-dimensional rectangular box in the parameter space. A
more flexible separating surface can be obtained if the variables are combined
in a polynomial instead. In this analysis a non-linear discriminant $D$ of
the form
\begin{eqnarray}
D ={\bf x^T}({\bf a}+ {\bf B x})
=a_1 x_1 +...+ a_n x_n + B_{11}x_1^2 +...+ B_{1n} x_1 x_n +...+ B_{nn}x_n^2
\end{eqnarray}
was defined~\cite{IDA}, where $x_i$ are the $n$ variables used, and $a_i$
and $B_{ij}$ are the corresponding weights. 
This can be written as
\begin{equation}
D = {\bf c^T  y} = c_1 y_1 +...+ c_m y_m,
\end{equation}
where {\bf c} contains all the weights $a_i$ and $B_{ij}$,
{\bf y} is the vector $({\bf x},{\bf x \cdot x^T})$, and $m=(n^2+3n)/2$.
To obtain a good signal-from-background separation, the variance
of $D$ should be as small as possible
while the separation in $D$ should be as large as possible.
Therefore the ratio
\begin{eqnarray}
 \rho = \frac{({\bf c} \Delta \mbox{\boldmath $\mu$})^2}{{\bf c^T V c}}
\end{eqnarray}
was maximized. Here ${\bf V}$ is the sum of the signal and background
variance matrices
\begin{eqnarray}
{\bf V = V}_{sig} + {\bf V}_{bkg}
\end{eqnarray}
and $\Delta \mbox{\boldmath $\mu$}$ is the difference between the signal and
background arithmetic means:
\begin{eqnarray}
\Delta \mbox{\boldmath $\mu$} = \mbox{\boldmath $\mu$}_{sig} -
\mbox{\boldmath $\mu$}_{bkg}
\end{eqnarray}
of all the $m$ variables in ${\bf y}$.
The discriminant was calculated
iteratively, and at each step of the iteration the
variables $x_i$ that gave the maximum background rejection were chosen. A
chosen signal efficiency was used to determine the $D$-cuts after each step.
The total number of variables used in the discriminant $D$, as well as the
number of iterative steps, and the efficiency of each step, can be varied in
the discriminant analysis.
A large number of variations of the combinations of variables used in the
discriminant, the number of steps, as well as the signal efficiencies, was
investigated before deciding on which parameter combination to use.

\section{Production of $\Xi^- \pi^+$ states}
\label{sec:Xic0rec}

Each $\Xi^-$ candidate in the mass range $1.30 < M(\Lambda \pi{_2^-}) < 1.34$
GeV/$c^2$, was combined with another pion, called $\pi_3$. It was required
that $\pi_3$ should have a charge opposite to the $\pi_1$ from
$\Lambda$ decay, and should not be tagged as a lepton.  
For all $(\Xi^- \pi{_3^+})$
combinations the invariant mass was calculated. 
If its value satisfied $2.2 < M(\Xi^- \pi{_3^+}) < 2.75$~GeV/$c^2$ 
(for $\Xi{_c^0}$ ), or 
$M(\Xi^- \pi{_3^+}) < 1.6$~GeV/$c^2$ (for $\Xi(1530)^0$), 
the iterative discriminant analysis described above was performed.

\subsection{$\Xi{_c^0} \rightarrow \Xi^- \pi^+$ selection}
\label{sec:Xic0sel}

\subsubsection{Separate simulation sets}
\label{sec:Xic0sets}

Two separate sets of simulated events were needed.
The training sample (called T{\footnotesize $(\Xi{_c^0})$} in the following) 
was used to find the weight vector ${\bf c}$ that gave a
maximum signal-from-background separation in $D$, 
while the analysis sample (called A{\footnotesize $(\Xi{_c^0})$}) was
used to determine the $\Xi{_c^0}$ efficiency (see Section~\ref{sec:fBRxic0}).
Both sets T{\footnotesize $(\Xi{_c^0})$} and A{\footnotesize $(\Xi{_c^0})$}
consisted of signal as well as
background events. 
The dedicated $\Xi{_c^0} \rightarrow \Xi^- \pi^+$
events described in Section~\ref{sec:eventsimu} were thus divided into
two sets: about 2/3 of the events were used 
in set T{\footnotesize $(\Xi{_c^0})$} and
the rest in set A{\footnotesize $(\Xi{_c^0})$}.
The simulated Z$\rightarrow \qq$ events
were used for the background in both cases, using only events without
any simulated $\Xi{_c^0}$'s. The signal in set T{\footnotesize $(\Xi{_c^0})$}
consisted of about 36000 events, while the background corresponded to
approximately $4.4\cdot10^6$ $q\bar{q}$ events.
After the constrained fit, and the above-mentioned selections, 
about 1900 signal and
75000 background events remained in set T{\footnotesize $(\Xi{_c^0})$} to be
used for the discriminant training. 
Set A{\footnotesize $(\Xi{_c^0})$} consisted of approximately 22000
$\Xi{_c^0} \rightarrow \Xi^- \pi^+$ events, plus
background events without any simulated $\Xi{_c^0}$ corresponding to
$5.3\cdot 10^6$ $q\bar{q}$ events. 
For each year, the number of background events 
in set A{\footnotesize $(\Xi{_c^0})$}
was weighted to correspond to the number of data events.
Each of the years 1992--1995 was trained separately.
The total numbers of events in the samples used for the $(\Xi{_c^0})$
analysis are shown in Table~\ref{nrofevts}.
\begin{table}[htb]
\begin{center}
\begin{tabular}{|l|r|r||l|r|}\hline
&  MC T{\footnotesize $(\Xi{_c^0})$}& MC A{\footnotesize $(\Xi{_c^0})$} &  & Data\\
\hline
\hline 
 Signal         &   35 678       & 22 133         & &    \\
 After cuts     &    1 883       &  1 268         & &    \\
\cline{1-3}   
\hline
 Background    & 4 426 483      &  5 329 144     & Total         & 3 498 492 \\
 After cuts    &   74 750       &   86 967       & After cuts    &    57 291 \\
\hline
\end{tabular}
\protect\caption{ \label{nrofevts}  {\em The numbers of signal and
background events in the simulation sets T{\footnotesize $(\Xi{_c^0})$} and
A{\footnotesize $(\Xi{_c^0})$}
(described in Section~\ref{sec:Xic0sets}), and in the data, both before and after the
multivertex fit and applied selections.  }}
\end{center}
\end{table}

\subsubsection{Parameter optimization}
\label{sec:paropt}

As mentioned above, a large number of parameter combinations was investigated
before a discriminant with two iterative steps, and seven variables,
was chosen. 
The chosen variables were the momentum $p(\Xi{_c^0})$, 
the fitted mass $M(\Xi^-)_{fit}$,
the angle between $\Xi^-$ and $\pi{_3^+}$,
the $b$-tagging probability, 
the $\Lambda$ decay length or the $\Xi^-$ decay length, 
the momentum $p(\Xi^-)$ or the $\Lambda$ mass, the impact
parameter of the $\pi{_2^-}$ or the momentum $p(\pi{_2^-})$,
with some variations among the different years.

\subsubsection{$\Xi{_c^0}$ production}
\label{sec:fBRxic0}

\begin{figure}[t]
\begin{center}
\mbox{\epsfig{figure=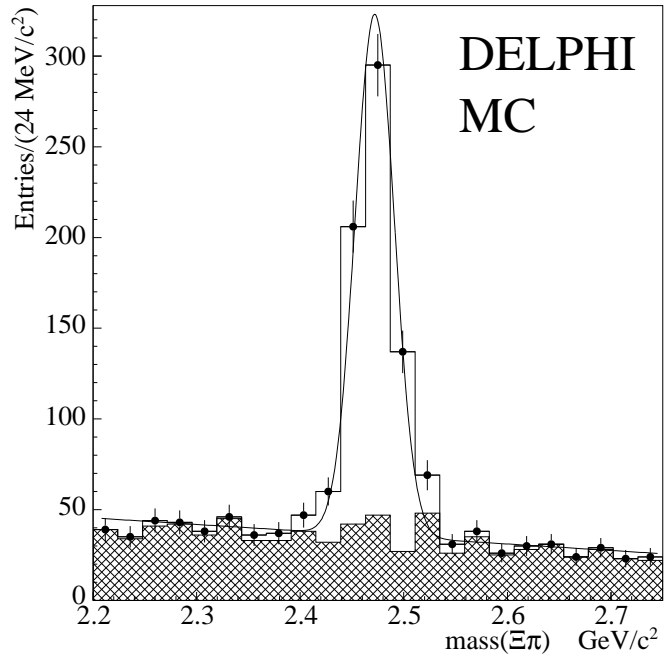,width=9cm}}
\protect\caption[mXicmc]{\label{mXicmc}  {\em 
The $\Lambda \pi{^-} \pi{^+}$
invariant-mass spectrum in the 1992--1995 simulation. 
The background (cross-hatched histogram) 
corresponds to\/ $5.3\cdot10^6$ Z$\rightarrow q\bar{q}$ simulated events. 
The signal (white histogram) consists of 
about 22000\/ $\Xi{_c^0} \rightarrow \Xi^- \pi^+$ simulated
events. 
The fitted curve, as described in Section~\ref{sec:fBRxic0}, uses 
a first-order polynomial to
parametrize the background and a Gaussian function for the signal.
}}
\end{center}
\end{figure}

Using the discriminants $D$ with the above mentioned variables 
on the simulation set A{\footnotesize $(\Xi{_c^0})$}, 
resulted in the ($\Lambda \pi^- \pi^+$) invariant-mass distribution shown in
Figure~\ref{mXicmc}. An extended unbinned maximum-likelihood fit was
performed.
The likelihood function used had the form:
\begin{eqnarray}
\log{L} = \log{(\mu_s f_s + \mu_b f_b)} -(\mu_s+\mu_b),
\end{eqnarray}
where $f_s$ is a Gaussian function used to parametrize the signal
and $f_b$, the probability density function for the background,
is a first order polynomial. 
The numbers of signal and background events were given by
$\mu_s$ and $\mu_b$, respectively. 
The fit gave $498 \pm 28$ events with a $\sigma$ width 
of 19$\pm$2~MeV/$c^2$ and mass $M(\Xi{_c^0}) =  2471\pm 1$~MeV/$c^2$, 
which can be compared with the generated mass of 2473.0~MeV/$c^2$. 
The number of true $\Xi{_c^0}$ events in
the peak was 494, and there was thus an excellent agreement.

\begin{figure}[h]
\begin{center}
\mbox{\epsfig{figure=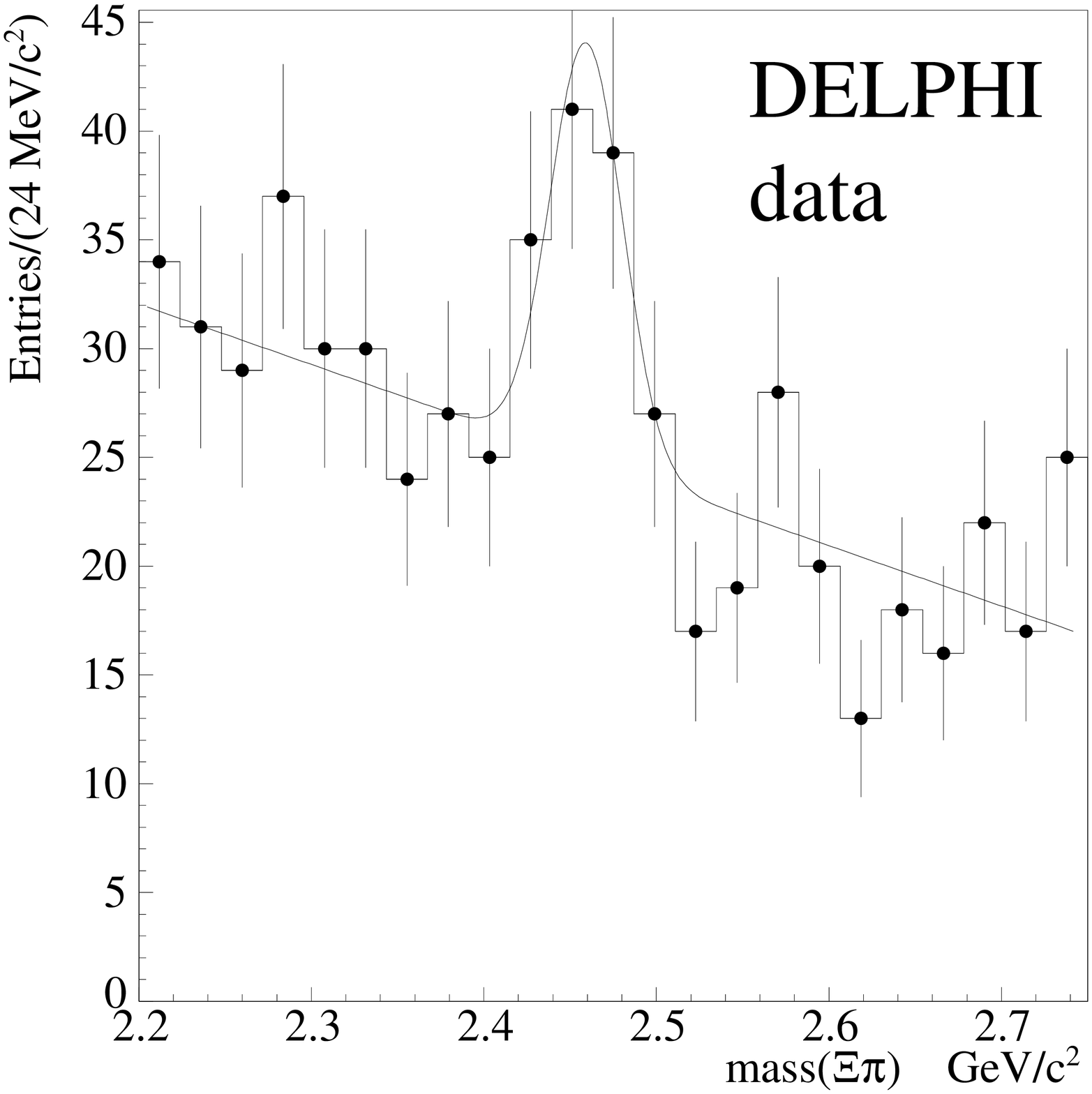,width=9cm}}
\protect\caption[mXicdat]{\label{mXicdat} {\em  
Same as Figure~\ref{mXicmc} for the 1992--1995 data sample, 
corresponding to\/ $3.5\cdot10^6$ hadronic Z decays.
}}
\end{center}
\end{figure}

The same discriminants were also used on the data, with the resulting 
mass distribution shown in
Figure~\ref{mXicdat}. In this case the fit gave 45$\pm$13 events
with a $\sigma$ width of 22$\pm$6~MeV/$c^2$. The mass from the fit was
2460$\pm$8~MeV/$c^2$, in agreement with the world average value
$M(\Xi{_c^0}) = 2471.8 \pm 1.4$~MeV/$c^2$~\cite{pdg}.

\begin{figure}[htb]
\begin{center}
\mbox{\epsfig{figure=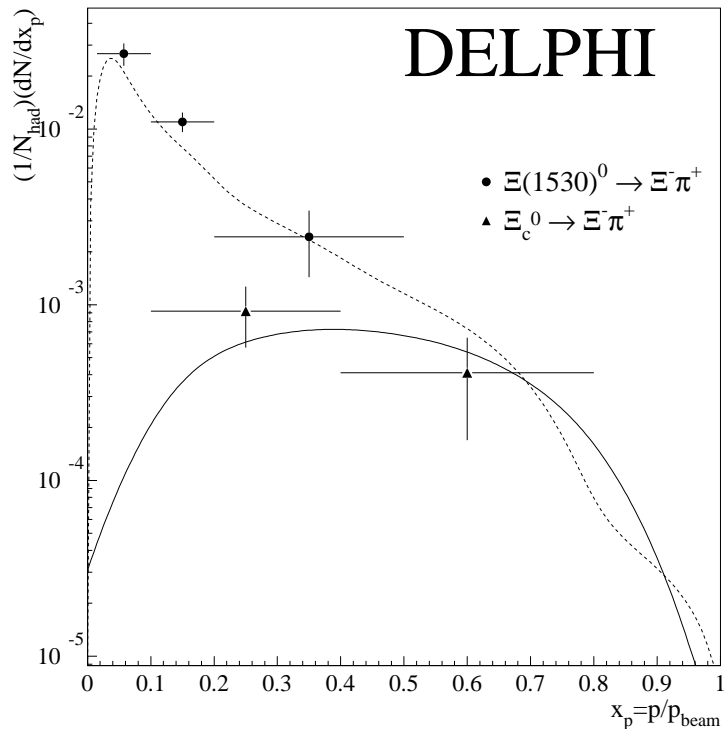,width=10cm}}
\protect\caption[dndxp]{\label{dndxp}  {\em The measured differential 
production rates for\/ $\Xi{_c^0}$ (plotted as triangles), and\/ $\Xi(1530)^0$ 
(circles).
For comparison the JETSET 7.4~\cite{JETSETfys} prediction is shown,
with a full line for\/ $\Xi{_c^0}$, and a dashed line for\/ $\Xi(1530)^0$. }}
\end{center}
\end{figure}

The reconstruction efficiency is highly momentum dependent, and
in order to avoid biases due to the $\Xi{_c^0}$ momentum distribution not
having been correctly described by the simulation,
the $\Xi{_c^0}$
rate was determined in two momentum bins, as shown in Table~\ref{dndxplist}
and Figure~\ref{dndxp}.
Integrating the observed distribution and using the JETSET 
generator~\cite{JETSETfys} 
to estimate the fraction of events outside
the observable momentum region (5.5\%), a total rate of 
\begin{eqnarray}
f_{\Xi{_c^0}} \times \mbox{BR}(\Xi{_c^0} \rightarrow \Xi^- \pi^+)=(4.7
\pm 1.4 (stat.)) \times 10^{-4}
\end{eqnarray}
per hadronic Z decay was obtained.
\begin{table}[htb]
\begin{center}
\begin{tabular}{|c|r|c|c|c|}\hline
Particle & $x_p$ interval & Efficiency & Nb. of particles & 
$\frac{1}{N_{had}}\frac{dN}{dx_p}$  \\
\hline
\hline
$\Xi{_c^0}\rightarrow \Xi^- \pi^+ $ 
             & 0.10--0.40 & (2.9$\pm$0.2)\% & 28.2$\pm$10.6 & $(9.2 \pm 3.5)\times 10^{-4}$\\
             & 0.40--0.80 & (1.9$\pm$0.2)\% & 11.0$\pm$6.5 & $(4.1 \pm 2.4)\times 10^{-4}$\\
\hline
\hline
$\Xi(1530)^0\rightarrow \Xi^- \pi^+ $
             & 0.015--0.10&(3.4$\pm$0.5)\%&271.1$\pm$39.3&$(2.7 \pm 0.4)\times 10^{-2}$\\
             & 0.10--0.20 &(7.6$\pm$0.9)\%&293.1$\pm$37.5&$(1.1 \pm 0.1)\times 10^{-2}$\\
             & 0.20--0.50 &(1.5$\pm$0.4)\%&37.3$\pm$15.8 &$(2.4 \pm 1.0)\times 10^{-3}$\\
\hline
\end{tabular}
\protect\caption{ \label{dndxplist}  {\em The $\Xi{_c^0}$ and\/
$\Xi(1530)^0$ differential production rates. The efficiency
and number of particles from the fit in each of the $x_p=p/p_{beam}$ 
intervals are also given.}}
\end{center}
\end{table}

\subsection{$\Xi(1530)^0$ production}
\label{sec:Xi1530sel}

As for the $\Xi{_c^0}$ analysis, an iterative discriminant analysis 
was applied for the \Xfem\ selection, and two separate simulation sets 
were used.
No special $\Xi(1530)^0$ simulation was needed, and the
$\qq$ simulated events were used for both signal and
background events. 

\begin{figure}[htb]
\begin{center}
\mbox{\epsfig{figure=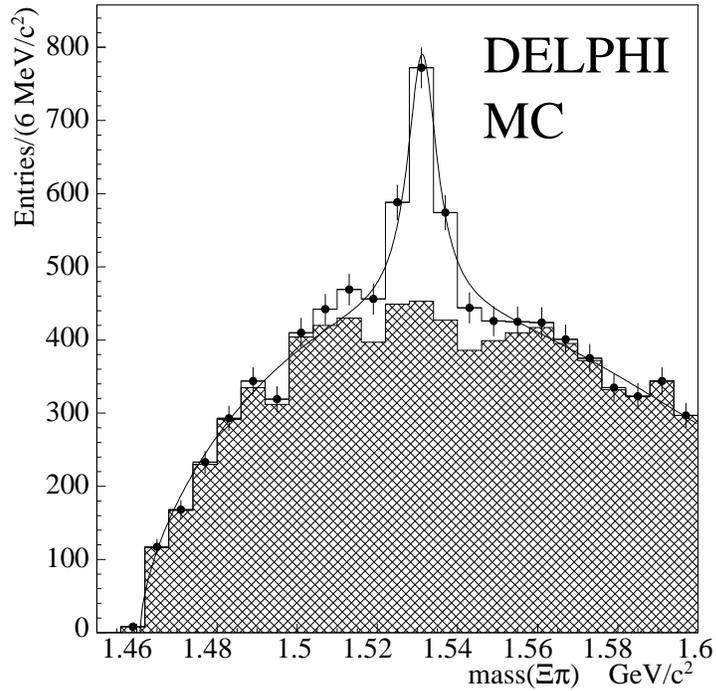,width=9.5cm}}
\protect\caption[mXi1530mc]{\label{mXi1530mc}  {\em  
The $\Lambda \pi{^-} \pi{^+}$ invariant-mass spectrum, 
using $5.3\cdot10^6$ simulated hadronic Z decays. The background
events are cross-hatched.
The fitted curve, as described in Section~\ref{sec:Xi1530sel}, 
uses a parametrized background and a Breit-Wigner function for the signal.
}}
\end{center}
\end{figure}

The \Xfem\ discriminant analysis was applied to the simulation and
resulted in the ($\Lambda \pi^- \pi^+$) invariant-mass distribution 
shown in Figure~\ref{mXi1530mc}.
The same maximum likelihood function as for the $\Xi{_c^0}$ was used, except
that the signal was described by a Breit-Wigner function, while the 
background was parametrized by~\cite{opal}
\begin{eqnarray}
\label{eqxi0bkg}
F(x) = (x - x_0)^{a} \times \exp{(b_0(x-x_0) + b_1(x-x_0)^2)},
\end{eqnarray}
where $x$ is the invariant mass measured in~GeV/$c^2$, $x_0$ is the 
kinematical limit of 1.4609~GeV/$c^2$, and $a$, $b_0$ and $b_1$ are 
free parameters. 
Since the signal peak is fairly close to the maximum of the background,
the fit easily became unstable. Therefore the width $\Gamma_0$ in the
Breit-Wigner was kept fixed
at the world average value of 9.1$\pm$0.5~MeV/$c^2$~\cite{pdg}.

Performing the fit on the simulation 
resulted in $844 \pm 71$ signal events, 
and a mass of $1531.2 \pm 0.5$~MeV/$c^2$, in good agreement with 
the number of \Xfem\ events in the sample (831), and the
generated mass of 1532.0~MeV/$c^2$.

\begin{figure}[htb]
\begin{center}
\mbox{\epsfig{figure=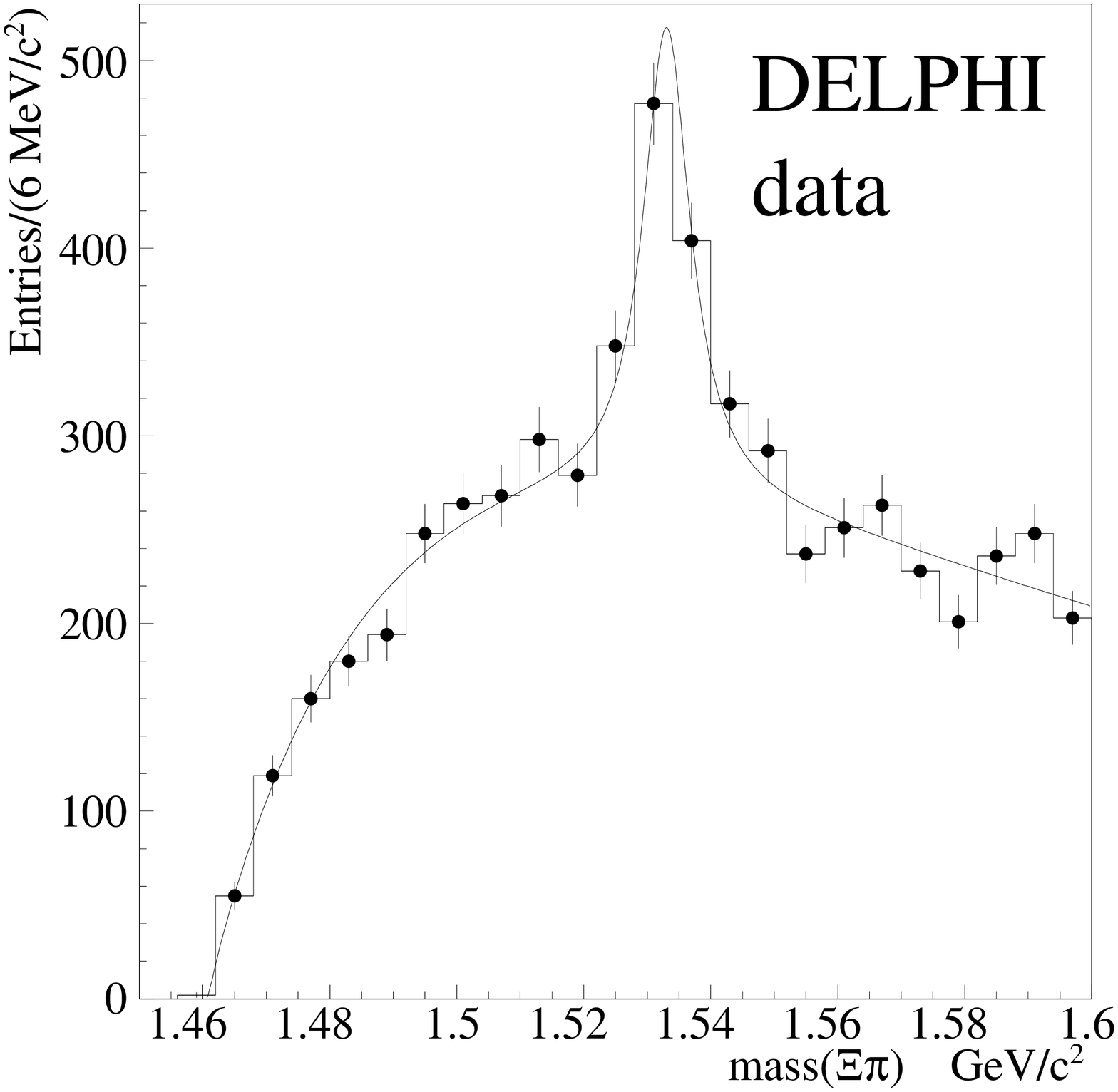,width=9.5cm}}
\protect\caption[mXi1530]{\label{mXi1530} {\em  
Same as Figure~\ref{mXi1530mc} for the 1992--1995 data sample, 
corresponding to\/ $3.5\cdot10^6$ hadronic Z decays.
}}
\end{center}
\end{figure}

 Using the same discriminant on the 1992--1995 data sample gave the 
invariant mass spectrum shown in Figure~\ref{mXi1530}. 
The unbinned maximum-likelihood fit gave $599 \pm 57$ signal
events in the peak, with a mass of $1533.0 \pm 0.5$~MeV/$c^2$, which can be
compared with the world average value 
$M(\Xi(1530)^0) = 1531.80 \pm 0.32$~MeV/$c^2$~\cite{pdg}.
The $\Xi(1530)^0$ production rate was evaluated in three momentum bins, as
shown in Table~\ref{dndxplist} and Figure~\ref{dndxp}. 
Integrating the observed distribution and using the JETSET 
generator~\cite{JETSETfys} to estimate the fraction of events outside
the observable momentum region (7.7\%), a total rate of 
\begin{eqnarray}
f_{\Xi(1530)^0} \times \mbox{BR}(\Xi(1530)^0 \rightarrow \Xi^- \pi^+)=
  (4.5\pm 0.5) \times 10^{-3}
\end{eqnarray}
per hadronic Z decay was obtained.

\subsection{Systematic uncertainties in production fractions}
\label{sec:syserr}

\subsubsection{The $\Xi{_c^0}$ baryon}
\label{sec:syserrxic}

Several different sources of systematic uncertainties were investigated.
The systematic uncertainty from the choice of the discriminant parameters was 
studied by varying the number of discriminant variables, 
the number of steps, and the signal efficiencies.
No significant variation beyond the expected statistical
fluctuation was found.

Due to the difference in the $ \Xi{_c^0}$ momentum distribution in
$c \rightarrow \Xi{_c^0}$ and $b \rightarrow c \rightarrow \Xi{_c^0}$  events, 
respectively, the reconstruction efficiencies in  $c\bar{c}$ and $b\bar{b}$
events can also differ. An important 
contribution to the systematic error was therefore 
the uncertainty in the 
different $c$ and $b$ efficiencies, as well as the relative production of
$\Xi{_c^0}$ in $c\bar{c}$ and $b\bar{b}$ events.
The number of observed $\Xi{_c^0}$ particles, $N_{seen}$, is given by
\begin{eqnarray}
\label{Nseen}
N_{seen} = N \times (R_c \cdot f_c \cdot \epsilon_c
+ R_b \cdot f_b \cdot \epsilon_b) \times \mbox{BR},
\end{eqnarray}
where $N$ is the total number of data events, 
$R_c$ and $R_b$ 
are the Z partial widths $\Gamma(c\bar{c})/\Gamma(hadrons)$ and 
$\Gamma(b\bar{b})/\Gamma(hadrons)$, $\epsilon_c$ and $\epsilon_b$ are the 
$\Xi{_c^0}$ reconstruction efficiencies for $c\bar{c}$ and $b\bar{b}$ events,
$f_c = f(c\bar{c} \rightarrow \Xi{_c^0})$ and
$f_b = f(b\bar{b} \rightarrow \Xi{_c^0})$ 
are the fractions of 
$c\bar{c}$ and $b\bar{b}$ quark pairs giving a $\Xi{_c^0}$, respectively 
(it has been assumed that $\Xi{_c^0}$ is only produced in $c\bar{c}$ and
$b\bar{b}$ events). Finally, $\mbox{BR}=\mbox{BR}(\Xi{_c^0} \rightarrow \Xi^-
\pi^+)$ is the branching fraction for the studied channel.

The reconstruction efficiencies for $c\bar{c}$ and $b\bar{b}$ events
were determined from the simulation and found to be 
$\epsilon_c = (1.4 \pm 0.2) \% $, and $\epsilon_b = (3.2 \pm 0.2) \%$. 
To estimate the relative weights of the $f_c$ and $f_b$ terms, both
simulation and data events were used. 
The $b$-purity of the $\Xi{_c^0}$ sample was enhanced with
a $b$-tag probability selection cut and
$f_b \times \mbox{BR}$ could then be found assuming that the
contribution from $c\bar{c}$ events was negligible.
The new $b$ efficiency $\epsilon_b^{(b)}$ was found from the simulation. 
The number of fitted $\Xi{_c^0}$ in data after this selection, 
$N_{seen}^{(b)}$,
was measured, and used to determine $f_b \times \mbox{BR}$ from:
\begin{eqnarray}
N_{seen}^{(b)} = N \times ( R_b \cdot f_b \cdot \epsilon_b^{(b)}) \times
\mbox{BR}.
\end{eqnarray}
It was found that $f_b\times\mbox{BR} = (1.3 \pm 0.6)\times 10^{-3}$,
essentially independent of the chosen $b$-tag cut. 

Using this value in equation~\ref{Nseen}, $f_c/f_b$ was found to be
compatible with 1.
For the study of the systematic uncertainty
$f_b\times\mbox{BR}$ was varied by one statistical standard deviation,
while $f_c/f_b$ was varied between 0.5 and 2.0 in equation~\ref{Nseen}.
The systematic uncertainty was thus estimated
to be $\pm 1.0 \times 10^{-4}$.

Other sources of systematics came from the contribution 
of the finite simulation statistics to the uncertainty
on the total $\Xi{_c^0}$ efficiency,
and the error rescaling done in the multivertex fit, 
giving a systematic uncertainty of  $\pm 0.3 \times 10^{-4}$ 
from each source.

As already mentioned, the JETSET model was used to estimate
the fraction of events in the unobserved momentum regions, which was found to
be 5.5\%. The comparison with JETSET (Figure~\ref{dndxp}) shows a good
agreement, thus the systematic uncertainty was estimated by varying
the value from the simulation by $\pm 50\%$, resulting in a systematic 
uncertainty contribution of $\pm 0.2\times 10^{-4}$.

All the above systematics are summarized in
Table~\ref{syscontr}. These  uncertainties were then added in quadrature,
which gave the final result:
\begin{eqnarray}
f_{\Xi{_c^0}} \times \mbox{BR}(\Xi{_c^0} \rightarrow \Xi^- \pi^+)=(4.7 \pm
1.4(stat.) \pm 1.1(syst.)) \times 10^{-4}.
\end{eqnarray}
\begin{table}[htb]
\begin{center}
\begin{tabular}{|c|c|c|}\hline
Source & $f \times $BR($\Xi{_c^0}$)& $f \times $BR($\Xi(1530)^0$)\\
\hline
\hline
Finite MC statistics & 0.3$\times 10^{-4}$&0.4$\times 10^{-3}$\\
MC extrapolation & 0.2$\times 10^{-4}$&0.2$\times 10^{-3}$\\
multivertex fit & 0.3$\times 10^{-4}$&0.4$\times 10^{-3}$\\
$b$, $c$ efficiencies & 1.0$\times 10^{-4}$&---\\
\hline
\hline
Total & 1.1$\times 10^{-4}$&0.6$\times 10^{-3}$\\
\hline
\end{tabular}
\protect\caption{ \label{syscontr}  {\em The different contributions to
the total systematic uncertainty for\/ $\Xi{_c^0}$ and\/
$\Xi(1530)^0$, as described in Section~\ref{sec:syserr}.
}}
\end{center}
\end{table}

\subsubsection{The $\Xi(1530)^0$ baryon}
\label{sec:sysxi1530}

The systematic uncertainties for $\Xi(1530)^0$ were studied in the same way 
as for the $\Xi{_c^0}$ events, except that since the $\Xi(1530)^0$ 
mainly comes from the fragmentation, these results were not affected 
by the flavour of the leading quark. 
All the different uncertainty contributions, summarized in Table~\ref{syscontr},
were then added in quadrature, and the final $\Xi(1530)^0$ result was
\begin{eqnarray}
f_{\Xi(1530)^0} \times \mbox{BR}(\Xi(1530)^0 \rightarrow \Xi^- \pi^+)=(4.5 \pm
0.5(stat.) \pm 0.6(syst.)) \times 10^{-3},
\end{eqnarray}
in agreement with the world average value of 
$(5.3 \pm 1.3)\times 10^{-3}$~\cite{pdg}
and the results of DELPHI~\cite{spyros} and OPAL~\cite{opal}.
The \Xfem\ result was used as a cross-check of the analysis method.

%
\section{Update of the $\Xi_b$ production rate and lifetime}
\label{sec:xibprod}

The strange $b$-baryon $\Xi_b$ was searched for 
in the semileptonic decay channel,
$\Xi_b \rightarrow \Xi^{-} \ell^{-} \bar{\nu} X$.
In the semileptonic decays of heavy hadrons the flavour of the
spectator system of the initial state is transmitted to the final
state. This property can be used to study $\Xi_b$ baryons from 
the observation of $\Xi^{\mp}$ production accompanied by a lepton
of the same sign.
The occurrence of  $\Xi^{\mp}- \ell^{\mp}$ pairs of same sign
(``right sign") 
is then compared to that of opposite sign pairs, $\Xi^{\mp}- \ell^{\pm}$
(``wrong sign").

\subsection{$\Xi^-$ and lepton reconstruction}
\label{sec:xilep}
The $\Xi^-$ was reconstructed using a constrained multivertex fit
as in the \Xzero\ analysis.
If the fit was successful, the $\Xi^-$ candidate was 
combined with a lepton candidate (electron or muon) 
within $1.0$ radian of the $\Xi^-$ momentum vector.
Since the expected production rate of the \Xb\ is
very small, loose selections were applied to the $\Xi^-$ and $\Lambda$
candidates.
The discriminant analysis method 
described above was used for the final \Xb\ selection.
Five variables were used in the discriminant;
the transverse momentum of the lepton with respect to the jet axis,
the invariant mass of the $\Xi^-$ and lepton, 
the combined momentum of the $\Xi^-$ and lepton, 
the number of charged particles in a 0.31 radian cone around
the lepton direction and the $\Xi^-$ variable,  
\mbox{$\xi = -$ln$x_p$}, where \mbox{$x_p = p_{\Xi}/p_{beam}$}.

\begin{figure}[htb]
\begin{center}
\mbox{\epsfig{figure=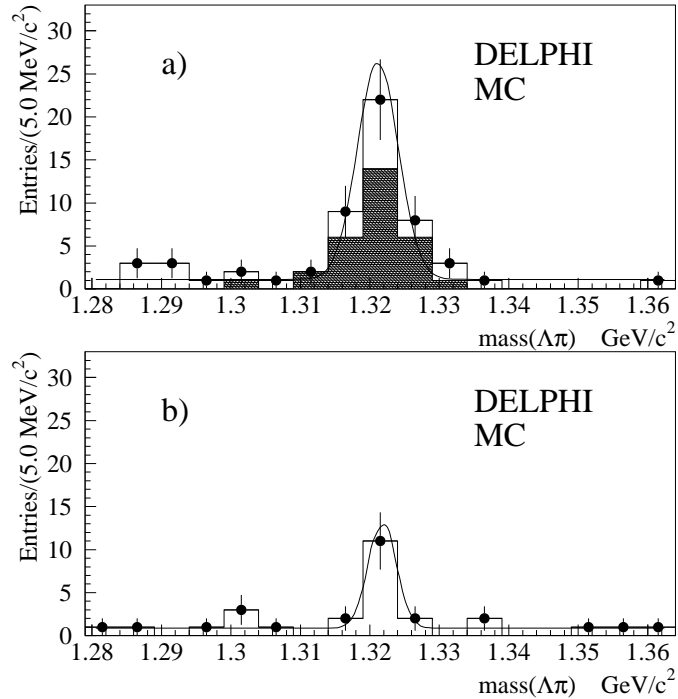,width=9cm}}
\protect\caption[$\Xi_b$ Results M.C.]{\label{rswsmcida} {\em  The 
$\Lambda \pi^-$ invariant-mass spectrum, using 
3.8 million simulated hadronic Z decays:
a) $\Xi^{\mp}- \ell^{\mp}$ right-sign pairs, 
b) $\Xi^{\mp}- \ell^{\pm}$ wrong-sign pairs.
The true \Xb\ simulated events are grey hatched.
The two fitted curves, as described in Section~\ref{sec:xilep}, each uses 
a constant value to parametrize the background 
and a Gaussian function for the $\Xi^-$ peak.
}}
\end{center}
\end{figure}

When applied to the Monte Carlo analysis sample,
consisting of 1/3 of the simulated events, 
the discriminant method gave the resulting $\Lambda \pi$
invariant-mass distributions
of Figure~\ref{rswsmcida}. 
Applying the extended unbinned maximum-likelihood fit 
described in Section~\ref{sec:fBRxic0} to the two distributions,
with a Gaussian function for the $\Xi^-$-peak 
and a constant value for the background, gave 34.2$\pm$5.9 right-sign events,
and 11.3$\pm$3.5 wrong-sign events,
when normalised to the size of the data sample.
The number of true $\Xi_b$ events in the right-sign $\Xi^-$ 
mass peak was $25.6$.
 Here the mass peak region was defined as the region
 with reconstructed $\Xi^-$ mass within $ \pm 10$ \MeVcc{}
 of the nominal $\Xi^-$ mass.

\begin{figure}[htb]
\begin{center}
\mbox{\epsfig{figure=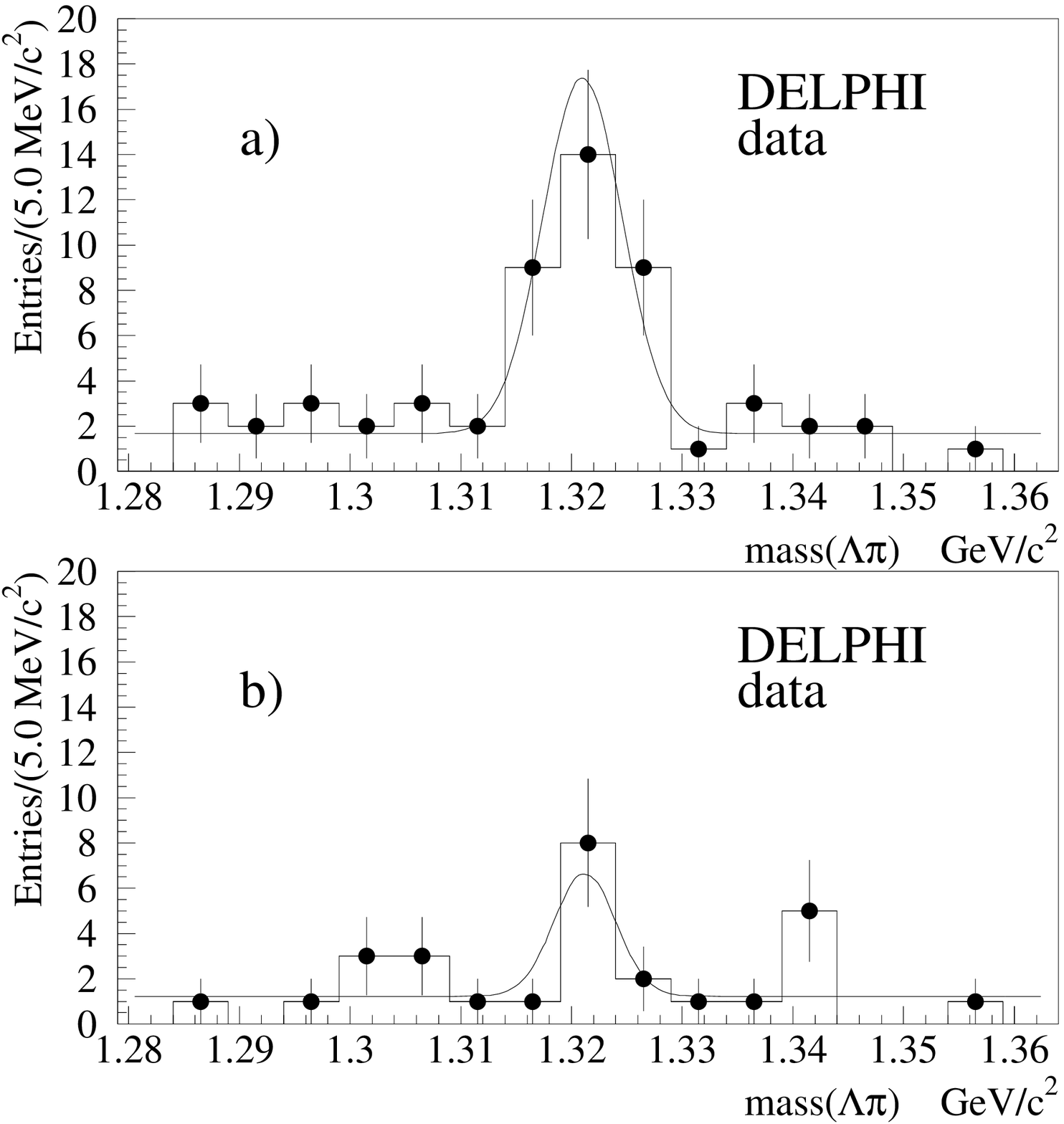,width=9cm}}
\protect\caption[$\Xi_b$ Results DATA]{\label{rswsdataida} {\em  
Same as Figure~\ref{rswsmcida} for the 1992--1995 data sample, 
corresponding to $3.5\cdot10^6$ hadronic Z decays.
}}
\end{center}
\end{figure}

The same analysis was applied to the full DELPHI
1992--1995 data sample and the resulting $\Lambda \pi$
invariant-mass distributions
are shown in Figure~\ref{rswsdataida}.
The unbinned maximum-likelihood fit to the two distributions,
gave a mean value of $1321.0 \pm 0.8$~\MeVcc{} 
for the right-sign distribution,
compatible with the nominal $\Xi^-$ mass 
($1321.31 \pm 0.13$~\MeVcc{}~\cite{pdg}).
The mean value from the fit to the right-sign distribution
was used as a fixed parameter in the fit to the wrong-sign distribution.
The maximum-likelihood fit resulted in 28.3$\pm$5.8 right-sign events,
and 7.6$\pm$3.3 wrong-sign events.

\subsection{$\Xi_b$ production rate}
\label{sec:xibrate}

 The number of background events in the right-sign 
$\Xi^-$ mass peak was estimated from that in the wrong-sign
mass peak, which resulted in \mbox{$20.7\pm 6.7$} \Xb\ events
found in the data. 
According to the simulation, the number of background events
is equal in the right- and wrong-sign $\Xi^-$ mass peaks. 
In the  Monte Carlo analysis sample the same procedure 
gave \mbox{$23.0\pm 6.8$} events 
to be compared with the number of true
$\Xi_b \rightarrow \Xi^{-}\ell^{-}X$ events
in the right-sign mass peak which was 25.6.

The total \Xb\ efficiency, calculated from the \Xb\ signal simulation
sample, was \mbox{$(2.3\pm0.1)$\%}.
Using the measured fraction of Z$\rightarrow \bb$
relative to all Z hadronic decays, 
$R_b$ = \mbox{($21.650 \pm 0.072)\%$}~\cite{pdg},
leads to a $\Xi_b$ production rate of:
\begin{eqnarray}
\mbox{BR}(b \rightarrow \Xi{_b})\times 
\mbox{BR}(\Xi{_b} \rightarrow \Xi^{-}\ell^{-}X) 
= (3.0\pm1.0(stat.))\times10^{-4}
\end{eqnarray}
per lepton species, 
averaged for electrons and muons. 

The dominating source of systematic uncertainty 
was the contribution of $\Lambda_b$ to the background.
The uncertainty in the background contribution of $\Lambda_b$
was estimated by varying
the amount of these events in the background by $\pm 20\%$ 
of the value in the simulation~\cite{pdg}.
This gave a shift of $\pm 0.20 \times 10^{-4}$ of the production rate.
 Other sources of systematic uncertainty
 were the finite simulation statistics,
 the error rescaling done in the multivertex fit,
 the Monte Carlo extrapolation of the events into the unobserved momentum
 regions (9.3\%) and a possible $\Xi_b$ polarisation.
All the above sources of systematic uncertainty are summarized in 
Table~\ref{systot}. 
The final result for the $\Xi_b$ production rate
was:
\begin{eqnarray}
\mbox{BR}(b \rightarrow \Xi{_b})\times 
\mbox{BR}(\Xi{_b} \rightarrow \Xi^{-}\ell^{-}X) 
= (3.0\pm1.0(stat.)\pm 0.3(syst.))\times10^{-4}
\end{eqnarray}
per lepton species, 
averaged for electrons and muons. 
This measurement of the production rate is in agreement with 
the previous measurements done by DELPHI~\cite{xibdelphi}
using a smaller data sample, and by
ALEPH~\cite{xibaleph}, see Table~\ref{bratios}.
In the previous DELPHI analysis~\cite{xibdelphi}, the background was estimated 
from the simulation while in this analysis
the background was estimated using the wrong-sign data sample.
\begin{table}[htb]
\begin{center}
\begin{tabular}{|c||c|}\hline
Source & Production rate variation \\
\hline
\hline
Finite MC stat.   & 0.10$\times 10^{-4}$ \\
Multivertex fit   & 0.15$\times 10^{-4}$  \\
MC extrapolation  & 0.15$\times 10^{-4}$ \\
$\Xi_b$ polarisation & 0.14$\times 10^{-4}$ \\
MC model of background   & 0.20$\times 10^{-4}$  \\
\hline
\hline
Total & 0.3$\times 10^{-4}$ \\
\hline
\end{tabular}
\protect\caption{ \label{systot} {\em The different contributions to
the total systematic uncertainty of the $\Xi_b$ production rate,
as described in Section~\ref{sec:xibrate}.}}
\end{center}
\end{table}
\begin{table}[htb]
\begin{center}
\begin{tabular}{|c||c|}\hline
Reference& Production rate   \\
\hline
\hline
 ALEPH~\cite{xibaleph}   & $(5.4\pm1.1(stat.)\pm 0.8(syst.))\times10^{-4}$ \\
 DELPHI~\cite{xibdelphi} & $(5.9\pm2.1(stat.)\pm 1.0(syst.))\times10^{-4}$ \\
\hline
\hline
 this analysis & $(3.0\pm1.0(stat.)\pm 0.3(syst.))\times10^{-4}$ \\
\hline
\end{tabular}
\protect\caption{ \label{bratios}  {\em Comparison between 
different results for the $\Xi_b$ production rate.
}}
\end{center}
\end{table}

\subsection{$\Xi_b$ lifetime}

The final sample of $\Xi^{-}\ell^{-}$ events 
is also used to measure the  $\Xi_b$ lifetime. 
The secondary vertex from the $\Xi_b$ semileptonic decay
is obtained by use of the BSAURUS package~\cite{bsaurus}.
The secondary vertex in the hemisphere is calculated in BSAURUS
using tracks that are likely to have originated from the decay chain
of a weakly decaying $b$-hadron state. The vertex fitting is done
in three dimensions, using the constraint of the direction of the
$b$-hadron candidate momentum vector, $p_{b}$.
The decay-length estimate in the $R\phi$ plane, is defined
as the distance between the primary- and secondary-vertex 
positions if the secondary-vertex fit was successful.
The three-dimensional decay length is obtained as 
$L = L_{R\phi} / \sin\theta$, 
where $\theta$ is the polar angle of the $b$-hadron candidate. 
The sign of the decay length is determined by the direction
of the $\Xi \ell $ momentum vector:
the distance is positive if the secondary vertex is found beyond
the primary vertex in this direction.
The resulting proper-time estimate for the $\Xi_b$ 
candidates is given by
$t = L \; m_{\Xi_b}/p_{b}$, where the $\Xi_b$ rest mass
is taken to be 5.8~\GeVcc. 

The lifetime was determined by an unbinned maximum-likelihood fit
to the proper time distribution.

The background was taken as the wrong-sign combinations plus 
the right-sign combinations outside the mass peak
but with a reconstructed $\Xi^-$ mass between 1.280~\GeVcc{} and
1.363~\GeVcc{}.
Both signal and background lifetimes were fitted
simultaneously. The purity in the right-sign mass peak
 was fixed in the fit and was taken 
from the simulation to be 0.67.

\begin{figure}[htb]
\begin{center}
\mbox{\epsfig{figure=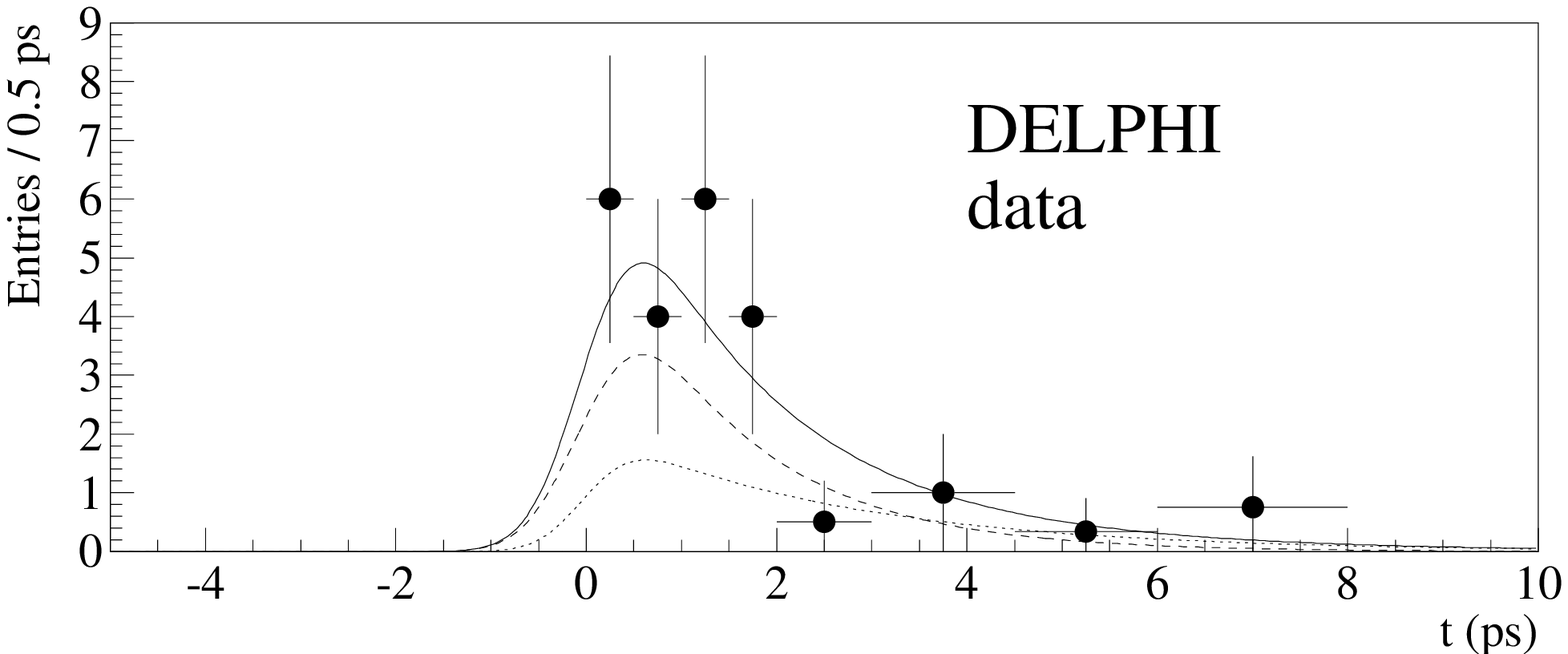,width=16cm}}
\protect\caption[$\Xi_b$ lifetime data]{\label{lifefit_data} {\em 
The result of the lifetime fit to the selected $\Xi_b$ events in
the data sample. The dotted curve is for the background, and the dashed
line corresponds to the signal. The full line is the total.}}
\end{center}
\end{figure}

In the 1992--1995 DELPHI data sample, 29 events in
the right-sign mass peak 
remained after the secondary vertex fit.
The statistical overlap of one event with the previous 
DELPHI lifetime measurement was removed. 
The result of the lifetime fit on the DELPHI data is presented in 
Figure~\ref{lifefit_data}, 
and gave
$\tau_{\Xi{_b}} = 1.45{^{+0.55}_{-0.43}}(stat.)$~ps.
The exact composition of the background, and the lifetimes of its
individual components, had no effect on the signal lifetime since
the background lifetime was fitted on the data.
Varying the fraction of $\Xi_b$ in the right-sign peak between 0.60 and 0.74
resulted in a variation of the fitted $\Xi_b$ lifetime of $\pm$0.10~ps.  
The proper time resolution was varied by $\pm$50\%, which resulted in a shift
of $\pm$0.07~ps.
The effect of a possible $\Xi_b$ polarisation was studied 
and found to be small.
Finally the $\Xi_b$ lifetime was measured to be
\begin{eqnarray}
\tau_{\Xi{_b}} = 1.45{^{+0.55}_{-0.43}} (stat.) \pm 0.13 (syst.)
\mbox{ ps}.
\end{eqnarray}
The measurement is in agreement with 
the previous measurements done by DELPHI~\cite{xibdelphi}
and by ALEPH~\cite{xibaleph}, see Table~\ref{life}.
\begin{table}[htb]
\begin{center}
\begin{tabular}{|c||c|}\hline
Reference & Lifetime (ps) \\
\hline
\hline
 ALEPH~\cite{xibaleph}   & $1.35{^{+0.37}_{-0.28}}(stat.){^{+0.15}_{-0.17}}(syst.)$ \\
 DELPHI~\cite{xibdelphi} & $1.5{^{+0.7}_{-0.4}}(stat.)\pm 0.3(syst.)$ \\
\hline
\hline
 this analysis &$1.45{^{+0.55}_{-0.43}}(stat.)\pm 0.13 (syst.)$ \\
\hline
\end{tabular}
\protect\caption{\label{life} {\em Comparison between 
different results for the $\Xi_b$ lifetime.
}}
\end{center}
\end{table}

The earlier DELPHI lifetime measurement~\cite{xibdelphi} used  
the data from 1991--1993 and a different method to 
reconstruct the $\Xi^-$-hyperon and the proper time,
and the background lifetime was estimated using simulation.
A combination of the two DELPHI lifetime measurements gives
\begin{eqnarray}
\tau_{\Xi{_b}} = 1.48{^{+0.40}_{-0.31}} (stat.) \pm 0.12 (syst.)
\mbox{ ps,}
\end{eqnarray}
using the method outlined in~\cite{average}.
The systematics are uncorrelated.

\section{Summary and conclusions}

The production rate per hadronic Z decay for the charmed baryon 
$\Xi{_c^0}$ has been measured for the first time:
$$
f_{\Xi{_c^0}} \times \mbox{BR}(\Xi{_c^0} \rightarrow \Xi^- \pi^+)=(4.7 \pm 
1.4(stat.) \pm 1.1(syst.)) \times 10^{-4}.
$$

As a cross-check, the $\Xi(1530)^0$ resonance was also reconstructed, and 
the corresponding production rate was found to be:
$$
f_{\Xi(1530)^0} \times \mbox{BR}(\Xi(1530)^0 \rightarrow \Xi^- \pi^+)=(4.5 \pm
0.5(stat.) \pm 0.6(syst.)) \times 10^{-3},
$$
in agreement with previous results~\cite{pdg}.

The beauty strange baryon $\Xi{_b}$ was searched for in the
semileptonic decay channel $\Xi{_b} \rightarrow \Xi^- \ell^- X$.
The product of the branching ratios in $b$ and $\Xi_b$ decays was
measured to be:
\begin{eqnarray*}
\mbox{BR}(b \rightarrow \Xi{_b})\times 
\mbox{BR}(\Xi{_b} \rightarrow \Xi^{-}\ell^{-}X) 
= (3.0\pm1.0(stat.)\pm 0.3(syst.))\times10^{-4}
\end{eqnarray*}
per lepton species, averaged for electrons and muons.

A measurement of the $\Xi_b$ lifetime gave:
$$\tau_{\Xi{_b}} = 1.45{^{+0.55}_{-0.43}} (stat.) \pm 0.13 (syst.)
\mbox{ ps},$$
in agreement with earlier results~\cite{xibdelphi,xibaleph}.
A combination of the two DELPHI lifetime measurements
gives
$$
\tau_{\Xi{_b}} = 1.48{^{+0.40}_{-0.31}} (stat.)\pm 0.12 (syst.)
\mbox{ ps.}
$$
%

\input{acknow.tex}

\newpage

\newpage

\newpage

\newpage

\newpage

\newpage

\end{document}

%% file: acknow.tex
\subsection*{Acknowledgements}
\vskip 3 mm
 We are greatly indebted to our technical 
collaborators, to the members of the CERN-SL Division for the excellent 
performance of the LEP collider, and to the funding agencies for their
support in building and operating the DELPHI detector.\\
We acknowledge in particular the support of \\
Austrian Federal Ministry of Education, Science and Culture,
GZ 616.364/2-III/2a/98, \\
FNRS--FWO, Flanders Institute to encourage scientific and technological 
research in the industry (IWT), Belgium,  \\
FINEP, CNPq, CAPES, FUJB and FAPERJ, Brazil, \\
Czech Ministry of Industry and Trade, GA CR 202/99/1362,\\
Commission of the European Communities (DG XII), \\
Direction des Sciences de la Mati$\grave{\mbox{\rm e}}$re, CEA, France, \\
Bundesministerium f$\ddot{\mbox{\rm u}}$r Bildung, Wissenschaft, Forschung 
und Technologie, Germany,\\
General Secretariat for Research and Technology, Greece, \\
National Science Foundation (NWO) and Foundation for Research on Matter (FOM),
The Netherlands, \\
Norwegian Research Council,  \\
State Committee for Scientific Research, Poland, SPUB-M/CERN/PO3/DZ296/2000,
SPUB-M/CERN/PO3/DZ297/2000, 2P03B 104 19 and 2P03B 69 23(2002-2004)\\
FCT - Funda\c{c}\~ao para a Ci\^encia e Tecnologia, Portugal, \\
Vedecka grantova agentura MS SR, Slovakia, Nr. 95/5195/134, \\
Ministry of Science and Technology of the Republic of Slovenia, \\
CICYT, Spain, AEN99-0950 and AEN99-0761,  \\
The Swedish Research Council,      \\
Particle Physics and Astronomy Research Council, UK, \\
Department of Energy, USA, DE-FG02-01ER41155. \\
EEC RTN contract HPRN-CT-00292-2002. \\
